
%
%
%
%
%
\newbox\leftpage \newdimen\fullhsize \newdimen\hstitle \newdimen\hsbody
\tolerance=1000\hfuzz=2pt
\def\printertype{ps: }
\def\qms{\def\printertype{qms: }
\ifx\answ\bigans\else\voffset=-.4truein\hoffset=.125truein\fi}
\def\bigans{b }
\message{ big or little (b/l)? }\read-1 to\answ
\ifx\answ\bigans\message{(This will come out unreduced.}
\magnification=1200\baselineskip=16pt plus 2pt minus 1pt
\hsbody=\hsize \hstitle=\hsize 
\else\message{(This will be reduced.} \let\lr=L
\magnification=1000\baselineskip=16pt plus 2pt minus 1pt
\voffset=-.31truein\vsize=7truein\hoffset=-.59truein
\hstitle=8truein\hsbody=4.75truein\fullhsize=10truein\hsize=\hsbody
\output={\ifnum\pageno=0 
  \shipout\vbox{\special{\printertype landscape}\makeheadline
    \hbox to \fullhsize{\hfill\pagebody\hfill}}\advancepageno
  \else
  \almostshipout{\leftline{\vbox{\pagebody\makefootline}}}\advancepageno 
  \fi}
\def\almostshipout#1{\if L\lr \count1=1 \message{[\the\count0.\the\count1]}
      \global\setbox\leftpage=#1 \global\let\lr=R
  \else \count1=2
    \shipout\vbox{\special{\printertype landscape}
      \hbox to\fullhsize{\box\leftpage\hfil#1}}  \global\let\lr=L\fi}
\fi
%
\catcode`\@=11 
\newcount\yearltd\yearltd=\year\advance\yearltd by -1900
 

\def\Title#1#2{\nopagenumbers\abstractfont\hsize=\hstitle\rightline{#1}%
\vskip 1in\centerline{\titlefont #2}\abstractfont\vskip .5in\pageno=0}
\def\Date#1{\vfill\leftline{#1}\tenpoint\supereject\global\hsize=\hsbody%
\footline={\hss\tenrm\folio\hss}}
%

\def\draftmode{\message{ DRAFTMODE }\def\draftdate{{\rm preliminary draft:
\number\month/\number\day/\number\yearltd\ \ \hourmin}}%
\headline={\hfil\draftdate}\writelabels\baselineskip=20pt plus 2pt minus 2pt
 {\count255=\time\divide\count255 by 60 \xdef\hourmin{\number\count255}
  \multiply\count255 by-60\advance\count255 by\time
  \xdef\hourmin{\hourmin:\ifnum\count255<10 0\fi\the\count255}}}
\def\nolabels{\def\wrlabel##1{}\def\eqlabel##1{}\def\reflabel##1{}}
\def\writelabels{\def\wrlabel##1{\leavevmode\vadjust
{\rlap{\smash{\line{{\escapechar=` \hfill\rlap{\hskip.03in\string##1}}}}}}}%
\def\eqlabel##1{{\escapechar=` \rlap{\hskip.09in\string##1}}}%
\def\reflabel##1{\noexpand\llap{\string\string\string##1}}}
\nolabels
%
\global\newcount\secno \global\secno=0
\global\newcount\meqno \global\meqno=1
\def\newsec#1{\global\advance\secno by1\message{(\the\secno. #1)}
\global\subsecno=0\xdef\secsym{\the\secno.}\global\meqno=1
\bigbreak\bigskip\noindent{\bf\the\secno. #1}\writetoca{{\secsym} {#1}}
\par\nobreak\medskip\nobreak}
\xdef\secsym{}
\global\newcount\subsecno \global\subsecno=0
\def\subsec#1{\global\advance\subsecno by1\message{(\secsym\the\subsecno. #1)}
\bigbreak\noindent{\it\secsym\the\subsecno. #1}\writetoca{\string\quad 
{\secsym\the\subsecno.} {#1}}\par\nobreak\medskip\nobreak}
\def\newsubsec#1{\global\advance\subsecno by1\message{(\secsym\the\subsecno. 
#1)}\bigbreak\noindent{\it\secsym\the\subsecno. #1}\writetoca{\string\quad 
{\secsym\the\subsecno.} {#1}}\par\nobreak\medskip\nobreak}
\def\appendix#1#2{\global\meqno=1\global\subsecno=0\xdef\secsym{\hbox{#1.}}
\bigbreak\bigskip\noindent{\bf Appendix #1. #2}\message{(#1. #2)}
\writetoca{Appendix {#1.} {#2}}\par\nobreak\medskip\nobreak}
%
%
\def\eqnn#1{\xdef #1{(\secsym\the\meqno)}\writedef{#1\leftbracket#1}%
\global\advance\meqno by1\wrlabel#1}
\def\eqna#1{\xdef #1##1{\hbox{$(\secsym\the\meqno##1)$}}
\writedef{#1\numbersign1\leftbracket#1{\numbersign1}}%
\global\advance\meqno by1\wrlabel{#1$\{\}$}}
\def\eqn#1#2{\xdef #1{(\secsym\the\meqno)}\writedef{#1\leftbracket#1}%
\global\advance\meqno by1$$#2\eqno#1\eqlabel#1$$}
%
\newskip\footskip\footskip14pt plus 1pt minus 1pt 
\def\f@@t{\baselineskip\footskip\bgroup\aftergroup\@foot\let\next}
\setbox\strutbox=\hbox{\vrule height9.5pt depth4.5pt width0pt}
\global\newcount\ftno \global\ftno=0
\def\foot{\global\advance\ftno by1\footnote{$^{\the\ftno}$}}
%
\newwrite\ftfile   
\def\footend{\def\foot{\global\advance\ftno by1\chardef\wfile=\ftfile
$^{\the\ftno}$\ifnum\ftno=1\immediate\openout\ftfile=foots.tmp\fi%
\immediate\write\ftfile{\noexpand\smallskip%
\noexpand\item{f\the\ftno:\ }\pctsign}\findarg}%
\def\footatend{\vfill\eject\immediate\closeout\ftfile{\parindent=20pt
\centerline{\bf Footnotes}\nobreak\bigskip\input foots.tmp }}}
\def\footatend{}
%
%
\global\newcount\refno \global\refno=1
\newwrite\rfile
\def\ref{[\the\refno]\nref}
\def\nref#1{\xdef#1{[\the\refno]}\writedef{#1\leftbracket#1}%
\ifnum\refno=1\immediate\openout\rfile=refs.tmp\fi
\global\advance\refno by1\chardef\wfile=\rfile\immediate
\write\rfile{\noexpand\item{#1\ }\reflabel{#1\hskip.31in}\pctsign}\findarg}
\def\findarg#1#{\begingroup\obeylines\newlinechar=`\^^M\pass@rg}
{\obeylines\gdef\pass@rg#1{\writ@line\relax #1^^M\hbox{}^^M}%
\gdef\writ@line#1^^M{\expandafter\toks0\expandafter{\striprel@x #1}%
\edef\next{\the\toks0}\ifx\next\em@rk\let\next=\endgroup\else\ifx\next\empty%
\else\immediate\write\wfile{\the\toks0}\fi\let\next=\writ@line\fi\next\relax}}
\def\striprel@x#1{} \def\em@rk{\hbox{}} 
\def\semi{;\hfil\break}
\def\addref#1{\immediate\write\rfile{\noexpand\item{}#1}} 
\def\listrefs{\footatend\vfill\eject\immediate\closeout\rfile\writestoppt
\baselineskip=14pt\centerline{{\bf References}}\bigskip{\frenchspacing%
\parindent=20pt\escapechar=` \input refs.tmp\vfill\eject}\nonfrenchspacing}
\def\startrefs#1{\immediate\openout\rfile=refs.tmp\refno=#1}
\def\xref{\expandafter\xr@f}\def\xr@f[#1]{#1}
\def\refs#1{[\r@fs #1{\hbox{}}]}
\def\r@fs#1{\edef\next{#1}\ifx\next\em@rk\def\next{}\else
\ifx\next#1\xref #1\else#1\fi\let\next=\r@fs\fi\next}
%

%
\newwrite\ffile\global\newcount\figno \global\figno=1
\def\fig{fig.~\the\figno\nfig}
\def\nfig#1{\xdef#1{fig.~\the\figno}%
\writedef{#1\leftbracket fig.\noexpand~\the\figno}%
\ifnum\figno=1\immediate\openout\ffile=figs.tmp\fi\chardef\wfile=\ffile%
\immediate\write\ffile{\noexpand\medskip\noexpand\item{Fig.\ \the\figno. }
\reflabel{#1\hskip.55in}\pctsign}\global\advance\figno by1\findarg}
\def\xfig{\expandafter\xf@g}\def\xf@g fig.\penalty\@M\ {}
\def\figs#1{figs.~\f@gs #1{\hbox{}}}
\def\f@gs#1{\edef\next{#1}\ifx\next\em@rk\def\next{}\else
\ifx\next#1\xfig #1\else#1\fi\let\next=\f@gs\fi\next}
\newwrite\lfile
{\escapechar-1\xdef\pctsign{\string\%}\xdef\leftbracket{\string\{}
\xdef\rightbracket{\string\}}\xdef\numbersign{\string\#}}

\def\writestop{\def\writestoppt{\immediate\write\lfile{\string\pageno%
\the\pageno\string\startrefs\leftbracket\the\refno\rightbracket%
\string\def\string\secsym\leftbracket\secsym\rightbracket%
\string\secno\the\secno\string\meqno\the\meqno}\immediate\closeout\lfile}}
\def\writestoppt{}\def\writedef#1{}
\def\seclab#1{\xdef #1{\the\secno}\writedef{#1\leftbracket#1}\wrlabel{#1=#1}}
\def\subseclab#1{\xdef #1{\secsym\the\subsecno}%
\writedef{#1\leftbracket#1}\wrlabel{#1=#1}}
\newwrite\tfile \def\writetoca#1{}
\def\leaderfill{\leaders\hbox to 1em{\hss.\hss}\hfill}
\def\writetoc{\immediate\openout\tfile=toc.tmp 
   \def\writetoca##1{{\edef\next{\write\tfile{\noindent ##1 
   \string\leaderfill {\noexpand\number\pageno} \par}}\next}}}

%
\catcode`\@=12 
%
\ifx\answ\bigans
\font\titlerm=cmr10 scaled\magstep3 \font\titlerms=cmr7 scaled\magstep3
\font\titlermss=cmr5 scaled\magstep3 \font\titlei=cmmi10 scaled\magstep3
\font\titleis=cmmi7 scaled\magstep3 \font\titleiss=cmmi5 scaled\magstep3
\font\titlesy=cmsy10 scaled\magstep3 \font\titlesys=cmsy7 scaled\magstep3
\font\titlesyss=cmsy5 scaled\magstep3 \font\titleit=cmti10 scaled\magstep3
\else
\font\titlerm=cmr10 scaled\magstep4 \font\titlerms=cmr7 scaled\magstep4
\font\titlermss=cmr5 scaled\magstep4 \font\titlei=cmmi10 scaled\magstep4
\font\titleis=cmmi7 scaled\magstep4 \font\titleiss=cmmi5 scaled\magstep4
\font\titlesy=cmsy10 scaled\magstep4 \font\titlesys=cmsy7 scaled\magstep4
\font\titlesyss=cmsy5 scaled\magstep4 \font\titleit=cmti10 scaled\magstep4
\font\absrm=cmr10 scaled\magstep1 \font\absrms=cmr7 scaled\magstep1
\font\absrmss=cmr5 scaled\magstep1 \font\absi=cmmi10 scaled\magstep1
\font\absis=cmmi7 scaled\magstep1 \font\absiss=cmmi5 scaled\magstep1
\font\abssy=cmsy10 scaled\magstep1 \font\abssys=cmsy7 scaled\magstep1
\font\abssyss=cmsy5 scaled\magstep1 \font\absbf=cmbx10 scaled\magstep1
\skewchar\absi='177 \skewchar\absis='177 \skewchar\absiss='177
\skewchar\abssy='60 \skewchar\abssys='60 \skewchar\abssyss='60
\fi
\skewchar\titlei='177 \skewchar\titleis='177 \skewchar\titleiss='177
\skewchar\titlesy='60 \skewchar\titlesys='60 \skewchar\titlesyss='60
\def\titlefont{\def\rm{\fam0\titlerm}
\textfont0=\titlerm \scriptfont0=\titlerms \scriptscriptfont0=\titlermss
\textfont1=\titlei \scriptfont1=\titleis \scriptscriptfont1=\titleiss
\textfont2=\titlesy \scriptfont2=\titlesys \scriptscriptfont2=\titlesyss
\textfont\itfam=\titleit \def\it{\fam\itfam\titleit} \rm}
\ifx\answ\bigans\def\abstractfont{\tenpoint}\else
\def\abstractfont{\def\rm{\fam0\absrm}
\textfont0=\absrm \scriptfont0=\absrms \scriptscriptfont0=\absrmss
\textfont1=\absi \scriptfont1=\absis \scriptscriptfont1=\absiss
\textfont2=\abssy \scriptfont2=\abssys \scriptscriptfont2=\abssyss
\textfont\itfam=\bigit \def\it{\fam\itfam\bigit}
\textfont\bffam=\absbf \def\bf{\fam\bffam\absbf} \rm} \fi
\def\tenpoint{\def\rm{\fam0\tenrm}
\textfont0=\tenrm \scriptfont0=\sevenrm \scriptscriptfont0=\fiverm
\textfont1=\teni  \scriptfont1=\seveni  \scriptscriptfont1=\fivei
\textfont2=\tensy \scriptfont2=\sevensy \scriptscriptfont2=\fivesy
\textfont\itfam=\tenit \def\it{\fam\itfam\tenit} 
\textfont\bffam=\tenbf \def\bf{\fam\bffam\tenbf} \rm}
%
%

\hyphenation{anom-aly anom-alies coun-ter-term coun-ter-terms}
\def\inv{^{\raise.15ex\hbox{${\scriptscriptstyle -}$}\kern-.05em 1}}

\def\Dsl{\,\raise.15ex\hbox{/}\mkern-13.5mu D} 
\def\dsl{\raise.15ex\hbox{/}\kern-.57em\partial}

\font\bigit=cmti10 scaled \magstep1
\def\lspace{\ifx\answ\bigans{}\else\qquad\fi}
\def\lbspace{\ifx\answ\bigans{}\else\hskip-.2in\fi} 
\def\boxeqn#1{\vcenter{\vbox{\hrule\hbox{\vrule\kern3pt\vbox{\kern3pt
	\hbox{${\displaystyle #1}$}\kern3pt}\kern3pt\vrule}\hrule}}}
\def\mbox#1#2{\vcenter{\hrule \hbox{\vrule height#2in
		\kern#1in \vrule} \hrule}}  
%

\def\darr#1{\raise1.5ex\hbox{$\leftrightarrow$}\mkern-16.5mu #1}

\def\roughly#1{\raise.3ex\hbox{$#1$\kern-.75em\lower1ex\hbox{$\sim$}}}


\long\def\optional#1{}
%
%

\def\dag{\dagger}     
\fontdimen16\tensy=2.7pt\fontdimen17\tensy=2.7pt 



%
%

\def\CL{{\cal L}}
\def\CM{{\cal M}}

%
%
\def\ts{\thinspace}
\def\ra{\rightarrow}

\def\Lra{\Longrightarrow}
\def\ra{\rightarrow}
\def\ol{\overline}

\def\W+-{W^\pm}
\def\Z0{Z^0}

\def\qbq{\ol{Q}Q}

\def\Mv{M_V}

\def\Ms+-{M^2_\pm}

\def\M+-{M_\pm}

\def\kslash{\raise.15ex\hbox{/}\kern-.57em k}
\def\pslash{\raise.15ex\hbox{/}\kern-.57em p}
%
%
%

\def\gev{{\rm GeV}}
\def\tev{{\rm TeV}}

\def\pb{{\rm pb}}
\def\fb{{\rm fb}}
\def\ecm{\sqrt{s}}

\def\shat{\hat s}

 
\def\simge{\mathrel{%
   \rlap{\raise 0.511ex \hbox{$>$}}{\lower 0.511ex \hbox{$\sim$}}}}
\def\simle{\mathrel{
   \rlap{\raise 0.511ex \hbox{$<$}}{\lower 0.511ex \hbox{$\sim$}}}}
 
\def\slashchar#1{\setbox0=\hbox{$#1$}           
   \dimen0=\wd0                                 
   \setbox1=\hbox{/} \dimen1=\wd1               
   \ifdim\dimen0>\dimen1                        
      \rlap{\hbox to \dimen0{\hfil/\hfil}}      
      #1                                        
   \else                                        
      \rlap{\hbox to \dimen1{\hfil$#1$\hfil}}   
      /                                         
   \fi}                                         %

\def\etmiss{\emiss_T}
%
%

%
%

%
%
%
%


\newwrite\ffile
\def\footend{ \def\foot{\global\advance\ftno by1\chardef\wfile=\ffile
$^{\the\ftno}$\ifnum\ftno=1\immediate\openout\ffile=foots.tmp\fi%
\immediate\write\ffile{\noexpand\smallskip%
\noexpand\item{f\the\ftno:\ }\pctsign}\findarg}
\def\listrefs{\vfill\eject\immediate\closeout\ffile\parindent=20pt
\centerline{{\bf Footnotes}}\bigskip\input foots.tmp
\vfill\eject\immediate\closeout\rfile\parindent=20pt
\baselineskip14pt\centerline{{\bf References}}\bigskip\frenchspacing%
\input refs.tmp\vfill\eject\nonfrenchspacing}
} 

\global\newcount\figno \global\figno=1
\newwrite\ffile
\def\pfig#1#2{Fig.~\the\figno\nfig#1{#2}}
\def\nfig#1#2{\xdef#1{Fig. \the\figno}%
\ifnum\figno=1\immediate\openout\ffile=figs.tmp\fi%
\immediate\write\ffile{\noexpand\item{\noexpand#1\ }#2}%
\global\advance\figno by1}

%
%
\def\tfig#1{Fig.~\the\figno\xdef#1{Fig. \the\figno}\global\advance\figno by1}

%
%
%
\def~{\ifmmode\phantom{0}\else\penalty10000\ \fi}
\def\lae{\raise-.5ex\vbox{\hbox{$\; <\;$}\vskip-2.9ex\hbox{$\; \sim\;$}}}
\def\gae{\raise-.5ex\vbox{\hbox{$\; >\;$}\vskip-2.9ex\hbox{$\; \sim\;$}}}


\def\fun#1#2{\lower3.6pt\vbox{\baselineskip0pt\lineskip.9pt
  \ialign{$\mathsurround=0pt#1\hfil##\hfil$\crcr#2\crcr\sim\crcr}}}

%
%

\overfullrule=0pt

\def\Lra{\Longrightarrow}
\def\pb{\rm pb}

\def\Mh{M_{\eta_T}}
\def\Mv{M_{V_8}}
\def\tbt{\ol t t}
\def\ttb{t \ol t}
\def\bbb{b \ol b}

\def\qbq{\ol q q}

\def\qqb{q \ol q}

\def\ppb{p \ol p}
\def\QbQ{\ol Q Q}

\def\stt{\sigma(\ttb)}
\def\Mtt{\CM_{t \ol t}}
\def\MMtt{\langle \Mtt \rangle}

\def\RMStt{{\langle \Mtt^2 \rangle^{1/2}}}
\def\Deltt{\Delta \Mtt}

\def\jets{{\rm jets}}

\def\et{E_T}

\def\et{E_T}
\def\etmiss{\slashchar{E}_T}
\def\cstar{\cos \theta^*}

\def\hl{10^{33} \ts {\rm cm}^{-2} \ts {\rm s}^{-1}}

\def\myfoot#1#2{{\baselineskip=14.4pt plus 0.3pt\footnote{#1}{#2}}}

\Title{\vbox{\baselineskip12pt\hbox{BUHEP--94--12}}}
{Top--Quark Production and Flavor Physics}
 
\bigskip
\centerline{Kenneth Lane\myfoot{$^{\dag }$}{lane@buphyc.bu.edu}}
\smallskip\centerline{Department of Physics, Boston University}
\centerline{590 Commonwealth Avenue, Boston, MA 02215}
\vskip .3in

\centerline{\bf Abstract}

Because of the top quark's very large mass, about 175~GeV, it now provides
the best window into flavor physics. Thus, pair--production of top quarks
at the Tevatron Collider is the best probe of this physics until the Large
Hadron Collider turns on in the next century. I will discuss aspects of the
mass and angular distributions that can be measured in $\ttb$ production
with the coming large data samples from the Tevatron and even larger ones
from the LHC.

\bigskip

\Date{6/94}

\vfil\eject 

\newsec{Introduction}

For more than a year, persistent rumors have been privately circulating
among particle physicists that signatures for the top quark have been found
in the CDF experiment at the Tevatron Collider. According to these rumors,
the ``best--fit'' top--quark mass is so large that the rate of the
signature events cannot be accounted for by ordinary QCD production of
$\ttb$
\ref\gieleref{J.~M.~Benlloch, K.~Sumorok, and W.~Giele,
``Possibilities of  Discovering a Heavy Top Quark in the Lepton-Multijet
Channel", FERMILAB--Pub--93/276--T~(1993) and references therein.}.
These rumors have now been confirmed in two papers from the CDF 
collaboration
\ref\cdfpr{F.~Abe, et al., The CDF Collaboration, {\it Evidence for
Top--Quark Production in $\ol p p$ Collisions at $\ecm = 1.8\,\tev$},
FERMILAB--PUB--94/097--E (1994), submitted to Physical Review~D.},
\ref\cdfprl{F.~Abe, et al., The CDF Collaboration, {\it Evidence for
Top--Quark Production in $\ol p p$ Collisions at $\ecm = 1.8\,\tev$},
FERMILAB--PUB--94/116--E (1994), submitted to Physical Review Letters.}
According to these papers, the top mass is $m_t = 174\pm 10 \ts ^{+13}_{-12}
\,\gev$.
The data in these papers are based on an integrated luminosity of
$19.3\,\pb^{-1}$. When combined with CDF's efficiencies and acceptances,
this yields the measured
cross section $\sigma(\ppb \ra \ttb) = 13.9 \ts^{+6.1}_{-4.8}\,\pb$
at $\sqrt{s} = 1800\,\gev$. For the central value of the measured top mass,
the predicted QCD cross section for this mass, including
next--to--leading--log and corrections and and soft--gluon resummation
\ref\qcdref{P.~Nason, S.~Dawson, and R.~K.~Ellis, Nucl.~Phys. {\bf B303}
(1988)~607; W.~Beenakker, H.~Kuijf, W.~L.~van~Neerven and
J.~Smith, Phys.~Rev. {\bf D40} (1989)~54.},
\ref\resum{E.~Laenen, J.~Smith and W.~L.~Van Neerven, Nucl.~Phys.~{\bf B369}
(1992)~543; {\it ibid}, FERMILAB--Pub--93/270--T.},
is $\stt = 5.10^{+0.73}_{-0.43}\,\pb$, 2.8~times smaller than the central value
of the measured cross section.

The experimental errors on the CDF measurements are large. But so is the
discrepancy with QCD, and this is tantalizing---evidence, we all hope, for
the long--sought breakdown of the standard model. In any case, it is clear
that the top quark is a wide open window into the world of flavor physics.
For example, at the top mass of 174~GeV, its Yukawa coupling to a standard
Higgs boson is $\Gamma_t = 2^{3/4}\ts G_F^{1/2}\ts m_t = 1.00$. In
Refs.~\ref\hillparke{C.~Hill and S.~Parke, Phys. Rev. {\bf D49}, 4454 (1994)}
and~\ref\eekleta{E.~Eichten and K.~Lane, Phys.~Lett.~{\bf B327}, 129 (1994).},
it was stressed that measurements of the $\ttb$ rate and associated
distributions at the Tevatron potentially provide the most powerful flavor
probes we now have. In particular, top--quark production can be
significantly modified from QCD expectations by the resonant production of
{\it colored}, flavor--sensitive particles with mass in the range
400--500~GeV. In Ref.~\hillparke, Hill and Parke proposed that 
color--octet vector mesons, $V_8$, associated with ``top color''
\ref\topcolor{C.~T.~Hill, Phys.~Lett.~{\bf B266}, 419 (1991)\semi
S.~P.~Martin, Phys.~Rev.~{\bf D45}, 4283 9a992); {\it ibid}, {\bf D46},
2197 (1992).}
are copious sources of $\ttb$.
In Ref.~\eekleta, Eichten and I stressed that the color--octet $\eta_T$
occurring in multiscale models of walking technicolor
\ref\etatrefs{E.~Farhi and L.~Susskind Phys.~Rev.~{\bf D20}
(1979)~3404\semi
S.~Dimopoulos, Nucl.~Phys.~{\bf B168} (1980)~69 \semi
T.~Appelquist and G.~Triantaphyllou, Phys.~Rev.~Lett.~{\bf
69} (1992)~2750 \semi
T.~Appelquist and J.~Terning, Yale and Boston University Preprint
YCTP-P21-93, BUHEP-93-23 (1993).}.
\ref\multi{K.~Lane and E.~Eichten, Phys.~Lett.~{\bf 222B} (1989)~274 \semi
K.~Lane and M~V.~Ramana, Phys.~Rev.~{\bf D44} (1991)~2678.},
\ref\wtc{B.~Holdom, Phys.~Rev.~{\bf D24} (1981)~1441;
 Phys.~Lett.~{\bf 150B} (1985)~301 \semi
T.~Appelquist, D.~Karabali and L.~C.~R. Wijewardhana,
Phys.~Rev.~Lett.~{\bf 57} (1986) 957~;
T.~Appelquist and L.~C.~R.~Wijewardhana, Phys.~Rev.~{\bf D36} (1987)~568
\semi  K.~Yamawaki, M.~Bando and K.~Matumoto, Phys.~Rev.~Lett.~{\bf 56},
(1986)~1335 \semi
T.~Akiba and T.~Yanagida, Phys.~Lett.~{\bf 169B} (1986)~432.}
is expected in just this mass range and easily could double the $\ttb$
rate expected from QCD. The top quark's mass is so large that, whatever
the nature of flavor physics, it is hard to believe that $m_t$ does not
offer some clue to it.

In this paper, we discuss two distributions that may reveal aspects of
flavor physics in $\ttb$ production---the invariant mass
distribution, $d\sigma/d\Mtt$, and the center--of--mass angular
distribution of the top quark, $d\sigma/d\cos\theta$. The magnitude and
shape of the invariant mass distribution will reveal whether
$\ttb$ production is standard or not, and whether resonances decaying to
$\ttb$ exist. We also point out that, for standard QCD production, the
$\Mtt$ distribution can provide an independent determination of the top
quark's mass. We apply this to the existing data~\cdfpr\ and find good
consistency with the reported mass. We also emphasize
the importance of measuring subsystem masses for testing alternative
top--production mechanisms.

The angular distribution of top quarks also reflects the underlying
production mechanism. Even though most of $\ttb$ production is near
threshold, the expectation that it is mainly s--wave can be overturned if
there are large parity--violating components in the $\qqb \ra \ttb$
process. We shall compare the angular distributions for standard and
nonstandard $\ttb$ production at the Tevatron and at the CERN Large 
Hadron Collider. We shall see that, because of the much larger $\tau = 
\shat/s$ of top--quark production at the Tevatron, experiments there
have a potential advantage over those at the LHC.

All these tests require much larger data sets than will be available in the
next year or two. To realize the full potential of this handle on flavor
physics, it is essential that the Tevatron experiments be able to collect
samples as large as 1--10~fb$^{-1}$. Such large data sets may even help
end what Mark Twain described as ``such wholesale returns of
conjecture out of such a trifling investment of fact.''

\newsec{Invariant Mass Distributions}

In QCD production of top--quark pairs, the mean and root--mean--square of
the $\ttb$ invariant mass distribution, $\MMtt$ and $\RMStt$, are nearly
linear functions of the top--quark mass
\ref\tdr{See, e.g., {\it GEM Technical Design Report}, Chapter~2; GEM
TN--93--262, SSCL--SR--1219; Submitted by the GEM Collaboration to the
Superconducting Super Collider Laboratory (April 30, 1993).}.
To understand why this is so, we show in Fig.~1 the $\Mtt$ distribution,
$d\sigma(\ppb \ra \ttb + X)/d \Mtt$, at $\sqrt{s} = 1800\,\gev$ for $m_t =$
100--220~GeV.\foot{These plots and all other calculations in this paper
were carried out using lowest--order QCD subprocess cross sections and the
EHLQ Set~1 parton distribution functions
\ref\ehlq{E.~Eichten, I.~Hinchliffe, K.~Lane and C.~Quigg, Rev.~Mod.~Phys.
{\bf 56}, 579 (1984).}.
We believe that our general conclusions will remain true when higher--order
corrections are included. Our $\ttb$ cross sections have been multiplied by
a factor of 1.6165. This makes our standard QCD rates as a function of
$m_t$ agree to within a per~cent with the central values quoted in
Ref.~\resum\ over the entire range of top masses of interest. Our numerical
results for the linear dependence of $\MMtt$ and $\RMStt$ on $m_t$ are
accurate so long as the higher--order corrections are well--represented by
a simple multiplicative factor. All our parton level calculations ignore
transverse motion of the $\ttb$ center--of--mass induced, e.g., by
initial--state radiation. While this effect is not large, it can and should
be taken into account in more detailed simulations.} For top quarks in this
mass range produced at the Tevatron, the cross section peaks reasonably
sharply at $\CM_{\rm max} \simeq 2.1 m_t + 10\,\gev$. At least for the
first few moments, then, we expect that
\eqn\mttmoment{
\langle \Mtt^n \rangle = {\int d\Mtt \ts\ts (d \sigma/d\Mtt) \ts \Mtt^n
\over {\int d\Mtt \ts\ts (d \sigma/d\Mtt) }} \ts,}
is a linear function of $m_t$. Using the lowest--order cross section, we
find that, for $100 \simle m_t \simle 200\,\gev$,
the first two moments are well--fit by the formulae
\eqn\mttvsmt{\eqalign{
\MMtt &= 50.0\,\gev + 2.24 \ts m_t \cr
\RMStt &= 58.4\,\gev + 2.23 \ts m_t \ts. \cr }}
In the range $m_t \simeq$ 140--180~GeV, the dispersion in $\Mtt$ expected
for standard QCD production is $\Deltt =$ 70--75~GeV.

In Ref.~\cdfpr, the top quark mass was determined from a sample of seven $W
\ra \ell \nu \ts + \ts 4 \ts \ts \jets$ events by making an overall
constrained best fit to the hypothesis $\ppb \ra \ttb + X$ followed by the
standard top decays $t \ra W^+ b$ with one $W$ decaying leptonically and
the other hadronically. At least one of the $b$--jets was tagged. The CDF
paper provides the momentum 4--vectors of all particles in the event before
and after the constrained fit. From these, the central values of kinematic
characteristics of the seven events may be determined. Table~1 lists the
best--fit top--quark masses determined by CDF together with the invariant
mass of the events before and after the constrained fit.\foot{Particle
4--vectors before the constrained fit do have various corrections---e.g.,
for the jet energy scale---made to them~\cdfpr. Only $\etmiss$ is provided
for the neutrino(s) in the before--fit 4--vectors. The biggest change in
the before-- and after--momenta occurs in $\etmiss$. We used the $W \ra
\ell \nu$ 4--momenta determined from the constrained fit in both cases.} We
used these $\Mtt$ to compute the mean and RMS. Both sets of 4--momenta gave
essentially identical results. Using 4--momenta from the constrained fit,
we found:
%
%
%
\eqn\after{\eqalign{
\MMtt &= 439\,\gev \quad \Lra \quad m_t = 173\,\gev \cr
\RMStt &= 443\,\gev \quad \Lra \quad m_t = 172\,\gev \cr
\Deltt &= 59.5\,\gev \ts. \cr}}

These results give some confidence that the measured central value of the
top--quark mass, 174~GeV, is accurate. For example, if $m_t = 160\,\gev$
(for which Ref.~\resum\ predicts $\stt = 8.2^{+1.3}_{-0.8}\,\pb$), we would
expect $\MMtt = 409\,\gev$ and $\RMStt = 415\,\gev$, both well below the
values determined above. Thus, if something is going to change in the CDF
results from the next large data sample, we expect it will be the cross
section---which would need to be two to three times smaller to agree with
the standard model.

Nonstandard explanations for the large cross section are not necessarily
disfavored by the good agreement between the central values of the measured
top mass and the top masses deduced in Eq.~\after. As an example of
nonstandard physics, we show in Figs.~2 and~3 two examples of $\ttb$
production rates with an $\eta_T$ resonance, one for $\Mh = 450\,\gev$, the
other for $\Mh = 475\,\gev$. The parameters are typical of those used in
the calculations of Ref.~\eekleta. They are listed in Table~2 along with
the production and kinematic characteristics of the two cases.\foot{The
$\eta_T$ contribution in the 475~GeV case was multiplied by 2.25 instead
of 1.62. This is consistent with the higher--order corrections to
$gg \ra \ttb$ processes predicted in Ref.~\resum.} The
top--quark masses inferred from the $\Mtt$ distributions are close to the
175~GeV input to these calculations. We note that the $\Mtt$ dispersion in
these examples is $\Deltt =$ 50--55~GeV, characteristic of
narrow--resonance production. Although this is closer to the measured
dispersion of 60~GeV (Eq.~\after) than the QCD expectation of about 75~GeV,
the statistics are too low for this agreement to be significant.

Subsystem invariant masses may be as interesting as the total invariant
mass. For example, in multiscale technicolor, it is possible that a
color--octet technirho is produced and decays as $\rho_T \ra W^\mp
\pi_T^\pm $, with $\pi_T^+ \ra t \ol b \ra W^+ b \ol b$, the {\it same}
final state as in $\ttb$ production~\multi. Searches for processes such as
these, using a constrained--fit procedure analogous to that employed by CDF
for the $\ttb$ hypothesis, should be carried out. All this will require
much more data from the Tevatron, probably $1\,\fb^{-1}$ or more. At the
expense of increasing backgrounds, larger data samples may be had by using
appropriately selected events {\it without} a tagged $b$--jet. This was
done already in Ref.~\cdfpr\ and was found to give an excess of events with
constrained--fit $m_t$ above 160~GeV.

To summarize: The invariant mass distributions that can be formed in
top--quark production provide incisive probes for distinguishing between
standard and non--standard mechanisms. In standard QCD, the mean and RMS of
$\Mtt$ provide an independent measure of $m_t$ which should agree with the
directly--measured mass. In QCD, the variance $\Deltt$ is expected to be
about 75~GeV. The total--system invariant mass can reveal the presence of
$\ttb$ resonances such as the $\eta_T$~\eekleta,\etatrefs\ and the
top--color vectors $V_8$ discussed by Hill and Parke~\hillparke,\topcolor.
Subsystem invariant masses can be studied to test for alternative
explanations of the top--production data. In this regard, we emphasize that
it is dangerous to use the standard QCD $\ttb$ production model to select
top--candidate events. For example, a resonance in $\ttb$ production can
distort the summed scalar--$\et$ and sphericity or aplanarity distributions
of candidate events from their QCD expectation.

\newsec{Angular Distributions}

The angular distribution of top quarks provides important information about
their production mechanism. In hadron colliders, the $\ttb$ pair is
produced by $q \ol q$ annihilation and $gg$ fusion. For pure QCD
production, the former process is dominant at the Tevatron---providing
about 90\% of the $\ttb$ rate. Gluon fusion dominates by the same amount at
much higher energy machines such as the LHC. Resonances such as the
$\eta_T$~\etatrefs, \eekleta\ and the ``top--color'' color--octet~$V_8$
vector boson \hillparke\ can change these proportions and the expected
angular distributions.\foot{Hill and Parke~\hillparke\ consider only $V_8$
bosons which have vector couplings to quarks. Such a $V_8$ will not induce
a significant change in the shape of the expected QCD angular distribution
at the Tevatron and LHC. For this reason, we consider below a $V_8$ which
couples only to one chirality of the quarks. The effect of this can be
dramatic.}

By Bose symmetry, the center--of--mass angular distribution in $gg \ra
\ttb$ is forward--backward symmetric. Although this is also true in
lowest--order QCD for $\qqb \ra \ttb$, there is no reason that it need be
so for non--standard production mechanisms. For example, if a $V_8$ exists
and couples only to left--handed quarks, the angular distribution in the
subprocess c.m. will be $(1 + \beta \cos \theta)^2$, where $\theta$ is the
angle between the incoming light quark and the outgoing top quark and
$\beta$ is the top quark's velocity.

%
%
In the study of top angular distributions, the Tevatron has a distinct
advantage.
In $\ppb \ra \ttb$ at $\sqrt{s} = 1800\,\gev$, the direction of the
incoming quark is the same as that of the proton practically all the time.
Thus, if we denote by $\theta^*$ the angle between the proton direction and
the top--quark direction in the subprocess~c.m., this angle is almost
always the same as $\theta$.\foot{The distinction between $t$ and $\ol t$
is based on the sign of the charged lepton in $W$--decay.}

In $pp$ collisions, the direction of the incoming quark can be inferred
with confidence only for events with high boost rapidity, $\eta_B$, or
large fractional subprocess energy, $\tau = \shat/s$. For large $\tau$, the
direction of the quark tends to be the same as the boost of the c.m., even
if $\eta_B$ is small \tdr. (In the case of $pp$ collisions, $\theta^*$ will
refer to the angle between the direction of the boost and that of the top
quark in the subprocess c.m.) Thus, angular information on top production
is doubly difficult to come by in $\ttb$ production at the LHC: The process
is dominated by gluon fusion, potentially obscuring interesting $\cos
\theta$ dependence, and $\tau$ is small, making it hard to distinguish
$\theta$ from $\pi - \theta$. As we shall see, the Tevatron's analyzing
power would be significantly greater if the luminosity of the Tevatron were
increased to $\hl$ or more and its detectors upgraded to handle (and
survive) this luminosity.\foot{The reader may have noticed that I did
not mention high--energy $e^+e^-$ colliders such as the 500~GeV or so NLC.
It is clear from the discussion here that lepton machines cast no light on
such strongly--coupled flavor physics aspects of $\ttb$ production as the
$\eta_T$ and $V_8$. The higher rates possible at hadron machines also
make them ideal for searches for new particles in top--quark decays.}

To illustrate the ability of high--luminosity hadron collider experiments
to distinguish among different production mechanisms, we compare the
$\cstar$ distributions for QCD, an $\eta_T$ of mass 450~GeV with isotropic
production, and a 475~GeV color--octet $V_8$ that couples only to
left--handed quarks. The angular distributions for the subprocesses $\qqb
\ra \ttb$ and $gg \ra \ttb$ in lowest--order QCD are:
\eqn\qcdang{\eqalign{
{d \hat \sigma(\qqb \ra \ttb) \over {d z}} =
{\pi \alpha_s^2 \beta \over {9 \shat}} \ts &\bigl(2 - \beta^2 + \beta^2
z^2\bigr) \ts,
\cr\cr
{d \hat \sigma(gg \ra \ttb) \over {d z}} =
{\pi \alpha_s^2 \beta \over {6 \shat}} \ts
&\biggl\{{1 + \beta^2 z^2 \over {1 - \beta^2 z^2}} -
{(1-\beta^2)^2 \ts (1 + \beta^2 z^2) \over{(1-\beta^2 z^2)^2}}
- \textstyle{{9\over{16}}} (1 + \beta^2 z^2) \cr 
&\ts\ts + {1-\beta^2 \over{1-\beta^2 z^2}} \ts (1 -
\textstyle{{1\over{8}}} \beta^2 + \textstyle{{3\over{8}}} \beta^2 z^2)
\biggr\} \ts, \cr}}
where $z = \cos\theta$ and $\beta = \sqrt{1 - 4 m_t^2/\shat}$. For $\shat
\gg 4 m_t^2$, these cross sections---especially the gluon fusion one---are
forward--backward peaked. But, at the modest $\shat$ at which QCD $\ttb$
production is large, the cross sections are fairly isotropic.

%
%
%

If there exists an $\eta_T$ with decay rates to gluon and quark pairs given
by~\eekleta\
\eqn\etarates{\eqalign{
\Gamma(\eta_T \ra gg) &= {5 \alpha_s^2 \ts N_{TC}^2 \ts \Mh^3  \over {384 \ts
\pi^3 \ts F_Q^2}} \ts, \cr\cr
\Gamma(\eta_T \ra \qbq) &= {C_q^2 \ts m_q^2 \ts \Mh \ts \beta_q \over {16 \pi
F_Q^2}} \ts,\cr}}
the gluon fusion cross section for $\ttb$ production has the following
additional terms:
\eqn\sigeta{\eqalign{
{d \hat \sigma(gg \ra \eta_T \ra \ttb) \over {d z}} =
&{\pi \over{4}} \ts {\Gamma(\eta_T \ra gg)\ts \Gamma(\eta_T \ra \tbt) \over
{(\shat - \Mh^2)^2 + \shat \ts \Gamma^2(\eta_T) }} \cr\cr
&\ts\ts + {5 \sqrt{2} \ts \alpha_s^2 \ts N_{TC}\ts C_t \ts m_t^2 \ts \beta
\over {768 \pi F_Q^2}} \ts {\shat - \Mh^2 \over
{(\shat - \Mh^2)^2 + \shat \ts \Gamma^2(\eta_T) }} \ts
\ts {1 - 2 \beta^2 z^2 \over {1 - \beta^2 z^2}} \ts. \cr}}
In these expressions, it is assumed that the $\eta_T$ is composed from a
single doublet of techniquarks $Q = (U,D)$ in the ${\bf N_{TC}}$
representation of $SU(N_{TC})$; $F_Q$ is the decay constant of technipions
in the $\QbQ$ sector; and $C_q$ is a dimensionless factor of $O(1)$ in the
Yukawa coupling of $\eta_T$ to $\qbq$. The second (interference) term in
the $\eta_T$ angular distribution is never very important, but we include
it for completeness.

The color--octet vector boson, $V_8$, is assumed to couple to $\qbq$
as follows:
\eqn\topvqq{
A(V^a_8(p,\lambda) \ra q(p_1) \ts \ol q(p_2)) =
g_s \ts \xi_q \ts \epsilon^\mu(p,\lambda) \ts
\ol u_q(p_1) \ts {\lambda_a \over{2}} \ts \gamma_\mu \ts \left({1-\gamma_5
\over{2}}\right)\ts v_q (p_2) \ts,}
where, following Ref.~\hillparke, $g_s$ is the QCD coupling and $\xi_t =
\xi_b = \pm 1/\xi_q$ ($q =u,d,c,s$). For this chiral coupling, the $\qqb
\ra \ttb$ angular distribution in Eq.~\qcdang\ is modified by the addition of
\eqn\sigveight{
{d \hat \sigma(\qqb \ra V_8 \ra \ttb) \over {d z}} =
{\pi \alpha_s^2 \beta \over {36 \shat}} \ts (1 + \beta z)^2
\ts \left\{\ts\biggl|1 + \xi_q \ts\xi_t \ts{\shat \over {\shat - \Mv^2 + i
\sqrt{\shat} \ts \Gamma(V_8)}}\biggr|^2 - 1\right\} \ts,}
where, ignoring the mass of all quarks except the top's, the $V_8$ width is
\eqn\widveight{
\Gamma(V_8) = {\alpha_s \Mv \over {12}}\ts
\biggl\{4\xi_q^2 + \xi_t^2 \left(1 + \beta (1-m_t^2/\Mv^2)\right)\biggr\} \ts.}

The $\cstar$ distributions we present below are an integral over $\ttb$
invariant mass of $d \sigma(p p^\mp \ra \ttb)/d\Mtt \ts d\cstar$. The
integration region is centered on the peak of the invariant mass
distribution and is approximately the width of the resonance. For the
$\eta_T$, we used $\Mh = 450\,\gev$, $N_{TC} = 5$, $F_Q = 30\,\gev$ and
$C_t = -1/3$. Then,
$$
\Gamma(\eta_T) \cong \Gamma(\eta_T \ra \ttb) +
\Gamma(\eta_T \ra gg) = 21\,\gev + 11\,\gev = 32\,\gev \ts,
$$
and $\sigma(\ppb \ra \ttb) \cong 14\,\pb$ at $\sqrt{s} = 1800\,\gev$. For
the $V_8$, we took $\Mv = 475\,\gev$ and $\xi_t = \sqrt{40/3}$ (see
\hillparke). Then,
$$
\Gamma(V_8) \cong \Gamma(V_8 \ra \bbb) + \Gamma(V_8 \ra \ttb) =
85\,\gev
$$
and $\sigma(\ppb \ra \ttb) \simeq 15\,\pb$ ($11\,\pb$) for $\xi_q =
-1/\xi_t$ ($\xi_q =  1/\xi_t$). Note that this $V_8$ model predicts a large
enhancement in $\qqb \ra \bbb$, providing a nice way to to test
it.\foot{Radiative corrections were approximated as above, by multiplying
the total lowest--order cross section by 1.62. This may overestimate the
$\qqb \ra \ttb$ contribution.}

The mass distribution for the 450~GeV $\eta_T$ was shown in Fig.~2 for the
Tevatron. Figures~4 and~5 show the $V_8$ mass distribution at the Tevatron
for $\xi_q = -1/\xi_t$ and $\xi_q = 1/\xi_t$, respectively. The effect of
the $V_8$'s interference with the $s$--channel gluon in $\qqb \ra \ttb$ and
its relation to the sign of $\xi_q\xi_t = \mp1$ are quite clear. The mean
and RMS $\ttb$ invariant masses for the two cases are, respectively:

\eqn\Mvminus{\eqalign{
\MMtt &= 440\,\gev \quad \Lra \quad m_t = 174\,\gev \cr
\RMStt &= 444\,\gev \quad \Lra \quad m_t = 173\,\gev \cr
\Deltt &= 53.4\,\gev \ts; \cr}}
\noindent and
\eqn\Mvplus{\eqalign{
\MMtt &= 482\,\gev \quad \Lra \quad m_t = 193\,\gev \cr
\RMStt &= 487\,\gev \quad \Lra \quad m_t = 192\,\gev \cr
\Deltt &= 66.6\,\gev \ts. \cr}}

The $\cstar$ distributions in $\ppb$ and $pp$ collisions, defined as
described above, are shown for the $\eta_T$ and $V_8$ models in Figs.~6--9
and their global features summarized in Table~3. In all cases, the
pseudorapidities of the $t$~and the $\ol t$~were required to be less
than~1.5 to allow for the mobility of their decay products and the finite
coverage of Tevatron and LHC detectors.\foot{This may be a bit unfair to
the LHC whose detectors ought to be much more hermetic and have somewhat
greater rapidity coverage for jets, electrons and muons than the Tevatron
detectors.} We discuss them in turn:

1.) Figure 6 shows the $\qqb \ra \ttb$, $gg \ra \ttb$ and $gg \ra \eta_T
\ra \ttb$ components of the top--production $\cstar$ distribution expected
at the Tevatron. The $\Mtt$ integration region is 430~to 470~GeV. The QCD
contribution is flat, the forward--backward peaking diminished by the
proximity of threshold. The $\eta_T$ contribution is also flat, of course,
and makes up about 80\% of the total cross section. The falloff near
$\cstar = \pm 0.90$ is due to the rapidity cut, $|\eta_{t, \ol t}| < 1.5$.
(For the fun of it, we computed the $\cstar$ distribution of the seven
$\ttb$~candidate events. The results, along with the c.m.~velocity~$\beta$,
are listed in Table~1. They form a perfectly flat distribution.) Table~3
lists the total $\ttb$~cross section as well as the cross sections
$\sigma_F$ for $\cstar > 0$ and $\sigma_B$ for $\cstar < 0$. The
forward-backward asymmetry is calculated as
\eqn\afb{
A_{FB} = {N_F - N_B \over{N_F + N_B}} =
{\sigma_F - \sigma_B \over{\sigma_F + \sigma_B}} \ts.}
The statistical error on $A_{FB}$ is
\eqn\Dafb{
(\Delta A_{FB})_{\rm stat} = 2 \sqrt{{N_F N_B \over {(N_F + N_B)^3}}} =
2 \sqrt{{\sigma_F \sigma_B \over {(\sigma_F + \sigma_B)^3 \ts
\epsilon_{t \ol t}\ts \int \CL dt}}} \ts,}
where $\epsilon_{t \ol t}$ is the overall efficiency, including branching
ratios, for identifying and reconstructing $\ttb$ events. For the CDF
experiment at the Tevatron, we can infer from Ref.~\cdfpr\ that
$\epsilon_{t \ol t}({\rm CDF}) \simeq$ 5--10 events$/(19\,\pb^{-1} \times
14\,\pb) =$ 2--4\%. We use $\epsilon_{t \ol t}({\rm TEV}) = 3\%$. It is
difficult to say what value of the efficiency is appropriate for LHC
experiments; detailed simulations are needed (see e.g., Ref.~\tdr).
We shall assume $\epsilon_{t \ol t}({\rm LHC}) = 5\%$, although it turns
out not to matter in the examples we consider.

The components of the $\cstar$ distribution expected at the LHC are shown
in Fig.~7. Because of the small $\tau$ values involved, the roles of gluon
fusion and $\qqb$ annihilation are reversed, with gluon fusion making up
about 90\% of the QCD rate. The enormous $\eta_T \ra \ttb$ rate is due to
the very large $gg$~luminosity at small~$\tau$~\ehlq. The central bowing of
the $\cstar$ distribution is due to the top--rapidity cut. At the LHC
energy, such large boost rapidities occur that events at large
c.m.~rapidity and $\cstar$ are depleted.

\medskip

2.) Figure 8 shows the components of the $\cstar$ distribution at the
Tevatron for the 475~GeV $V_8$ model coupling to left--handed quarks with
relative strengths $\xi_t = -1/\xi_q = \sqrt{40/3}$. The $\Mtt$ integration
region is 400--500~GeV. The effect of the chiral coupling is evident,
though somewhat diminished by the $\eta_{t, \ol t}$~cut. The
forward--backward asymmetry of~0.30 could be measured at the $4\sigma$
(statistical) level with an integrated luminosity of $1\,\fb^{-1}$. For
this luminosity, the statistical errors on $d\sigma/d\cstar$ in six bins
0.30~units wide would range from 30\% down to 15\%. This example shows how
useful it would be if the Tevatron luminosity could be upgraded to~$\hl$.

The $\cstar$ distributions expected at the LHC for this $V_8$ are shown in
Fig.~9. In this example, the contribution of the $V_8$ is about 20\% of the
total and it is polluted by the $q \leftrightarrow \ol q$ ambiguity, so
that the rise in the cross section with $\cstar$ is invisible. The
asymmetry is only 1\%. This illustrates the dominance of $gg$ processes and
the uncertainty in determining the quark direction at small~$\tau$ in a
high--energy $pp$~collider that we mentioned earlier. Essentially similar
results were obtained for the $\xi_t = 1/\xi_q$ case (see Table~3).

The pronounced central peaking in Fig.~9 is an artifact of the rather tight
$\eta_{t, \ol t}$~cut. It goes away for a looser cut, as seen in Fig.~10
for $\vert \eta_{t, \ol t} \vert < 2.5$. The asymmetry in $\cstar$ is still
invisible, however, and the error in $A_{FB} = 0.025$ probably would be
dominated by systematic effects. We also found that there is nothing to be
gained at the LHC by limiting the $\Mtt$ integration region to a narrow
band about $\Mv$ or by selecting events produced at large boost rapidity.

To summarize this section: The dominance of $\qqb$ annihilation in
top--quark production processes at the Tevatron collider gives it an
advantage over the LHC for studying angular distributions. However,
measurements of these distributions would benefit greatly from a
significant upgrade of the collider and its detectors so that data samples
of $O(10\,\fb^{-1})$ can be collected. The studies carried out here have
all been at the most naive parton level; it is hoped that detailed,
detector--specific simulations will be undertaken in the not--too--distant
future.

I am indebted to Elizabeth Simmons for a critical reading of the manuscript
and to Estia Eichten, John Huth and John Terning for many helpful
conversations. This research was supported in part by the Department of
Energy under Contract~No.~DE--FG02--91ER40676 and by the Texas National
Research Laboratory Commission under Grant~No.~RGFY93-278.

\footatend\vfill\eject\immediate\closeout\rfile\writestoppt
\baselineskip=14pt\centerline{{\bf References}}\bigskip{\frenchspacing%
\parindent=20pt\escapechar=` \input refs.tmp\vfill\eject}\nonfrenchspacing

\centerline{\vbox{\offinterlineskip
\hrule\hrule
\halign{&\vrule#&
  \strut\quad#\hfil\quad\cr
height4pt&\omit&&\omit&&\omit&&\omit&&\omit&&\omit&&\omit&\cr
&\hfill Run---Event \hfill&&\hfill$m_t$ \hfill&&\hfill
 $\Mtt$(before fit)\hfill&&\hfill
$\Mtt$(after fit) \hfill&&\hfill $\beta$(after fit)\hfill&& 
\hfill $\cstar$\hfill &\cr
height4pt&\omit&&\omit&&\omit&&\omit&&\omit&&\omit&\cr
\noalign{\hrule}
height4pt&\omit&&\omit&&\omit&&\omit&&\omit&&\omit&&\omit&\cr
&40758--44414&&\hfill$172\pm 11$\hfill&&\hfill$523$\hfill
&&\hfill$526$\hfill&&
\hfill$0.757$\hfill&&\hfill $0.404$\hfill&\cr
height4pt&\omit&&\omit&&\omit&&\omit&&\omit&&\omit&&\omit&\cr
&43096--47223&&\hfill$166\pm 11$\hfill&&\hfill$533$\hfill
&&\hfill$511$\hfill&&
\hfill$0.760$\hfill&&\hfill $0.820$\hfill&\cr
height4pt&\omit&&\omit&&\omit&&\omit&&\omit&&\omit&&\omit&\cr
&43351--266423&&\hfill$158\pm 18$\hfill&&\hfill$440$\hfill
&&\hfill$460$\hfill&&
\hfill$0.727$\hfill&&\hfill $0.512$\hfill&\cr
height4pt&\omit&&\omit&&\omit&&\omit&&\omit&&\omit&&\omit&\cr
&45610--139604&&\hfill$180\pm 9$\hfill&&\hfill$338$\hfill
&&\hfill$366$\hfill&&
\hfill$0.180$\hfill&&\hfill $-0.0011$\hfill&\cr
height4pt&\omit&&\omit&&\omit&&\omit&&\omit&&\omit&&\omit&\cr
&45705--54765&&\hfill$188\pm 19$\hfill&&\hfill$440$\hfill
&&\hfill$431$\hfill&&
\hfill$0.489$\hfill&&\hfill $-0.348$\hfill&\cr
height4pt&\omit&&\omit&&\omit&&\omit&&\omit&&\omit&&\omit&\cr
&45879--123158&&\hfill$169\pm 10$\hfill&&\hfill$411$\hfill
&&\hfill$412$\hfill&&
\hfill$0.572$\hfill&&\hfill $-0.767$\hfill&\cr
height4pt&\omit&&\omit&&\omit&&\omit&&\omit&&\omit&&\omit&\cr
&45880--31838&&\hfill$132\pm 8$\hfill&&\hfill$384$\hfill
&&\hfill$365$\hfill&&
\hfill$0.691$\hfill&&\hfill $-0.682$\hfill&\cr
height4pt&\omit&&\omit&&\omit&&\omit&&\omit&&\omit&\cr}
\hrule\hrule}}
\medskip
\centerline{TABLE 1: Best fit top--quark masses (Ref.~\cdfpr) and
kinematic characteristics of the $\ttb$ candidate events.}
\medskip
\noindent{Masses are in GeV. Transverse motion of the subprocess c.m. was
neglected in determining the top--quark velocity $\beta$ and scattering
angle $\theta^*$. }

\bigskip\bigskip\bigskip\bigskip

\centerline{\vbox{\offinterlineskip
\hrule\hrule
\halign{&\vrule#&
  \strut\quad#\hfil\quad\cr
height4pt&\omit&&\omit&&\omit&&\omit&&\omit&&\omit&&\omit&\cr
&\hfill $\Mh$ \hfill&&\hfill $\sigma(\ttb)$ \hfill&&\hfill
$\sigma_{\eta_T}(\ttb)$\hfill&&\hfill
$\MMtt$  \hfill&&\hfill $m_t\ts(\MMtt)$\hfill&&\hfill $\RMStt$\hfill&&
\hfill $m_t\ts(\RMStt)$ \hfill &\cr
height4pt&\omit&&\omit&&\omit&&\omit&&\omit&&\omit&&\omit&\cr
\noalign{\hrule}
height4pt&\omit&&\omit&&\omit&&\omit&&\omit&&\omit&&\omit&\cr
&$450$&&\hfill$13.5$\hfill&&\hfill$8.53$\hfill
&&\hfill$432$\hfill&&
\hfill$171$\hfill&&\hfill $435$\hfill&&\hfill $169$\hfill&\cr
height4pt&\omit&&\omit&&\omit&&\omit&&\omit&&\omit&&\omit&\cr
&$475$&&\hfill$13.9$\hfill&&\hfill$8.95$\hfill
&&\hfill$442$\hfill&&
\hfill$175$\hfill&&\hfill $445$\hfill&&\hfill $173$\hfill&\cr
height4pt&\omit&&\omit&&\omit&&\omit&&\omit&&\omit&&\omit&\cr}
\hrule\hrule}}
\medskip

\centerline{TABLE 2: $\ppb \ra \eta_T \ra \ttb$ cross sections (in pb) and
their kinematic characteristics.}

\medskip

\noindent {In the notation of Ref.~\eekleta, $N_{TC} = 5$, $F_Q =
30\,\gev$, and the coupling of the $\eta_T$ to $\ttb$ is $C_t = -1/3$. To
account for QCD radiative corrections, cross sections were multiplied
by~1.62. For $\Mh = 475\,\gev$, the $\eta_T$--contribution was multiplied
by~2.25. The QCD cross section is $4.96\,\pb$~\resum. The dispersion in
$\Mtt$ is $\Deltt = 52\,\gev$ for $\Mh = 450\,\gev$ and 55~GeV for $\Mh =
475\,\gev$.}

\vfil\eject

\centerline{\vbox{\offinterlineskip
\hrule\hrule
\halign{&\vrule#&
  \strut\quad#\hfil\quad\cr
height4pt&\omit&&\omit&&\omit&&\omit&&\omit&&\omit&&\omit&\cr
&\hfill Model \hfill&&\hfill $\Mtt$ range \hfill&&\hfill
 Collider\hfill&&\hfill
$\stt$ \hfill&&\hfill $\sigma_F$ \hfill&& 
\hfill $\sigma_B$ \hfill&& \hfill $A_{FB}$ \hfill &\cr
height4pt&\omit&&\omit&&\omit&&\omit&&\omit&&\omit&&\omit&\cr
\noalign{\hrule}
height4pt&\omit&&\omit&&\omit&&\omit&&\omit&&\omit&&\omit&\cr
&\hfill $\eta_T$ \hfill&&\hfill $430-470$ \hfill&&\hfill TEV\hfill
&&\hfill$4.03$\hfill&&
\hfill$2.01$\hfill&&\hfill $2.01$\hfill&&\hfill 0$$ \hfill &\cr
height4pt&\omit&&\omit&&\omit&&\omit&&\omit&&\omit&&\omit&\cr
&\hfill $\eta_T$ \hfill&&\hfill $430-470$ \hfill&&\hfill LHC\hfill
&&\hfill$2800$\hfill&&
\hfill$1400$\hfill&&\hfill $1400$\hfill&&\hfill $0$ \hfill &\cr
height4pt&\omit&&\omit&&\omit&&\omit&&\omit&&\omit&&\omit&\cr
&$V_8$ $(\xi_q = - 1/\xi_t$)&&\hfill $400-500$ \hfill&&\hfill TEV \hfill
&&\hfill$6.43$\hfill&&
\hfill$4.20$\hfill&&\hfill $2.24$\hfill&&\hfill $0.30$ \hfill &\cr
height4pt&\omit&&\omit&&\omit&&\omit&&\omit&&\omit&&\omit&\cr
&$V_8$ $(\xi_q = - 1/\xi_t$)&&\hfill $400-500$ \hfill&&\hfill LHC\hfill
&&\hfill$175$\hfill&&
\hfill$88$\hfill&&\hfill $86$\hfill&&\hfill $0.010$ \hfill &\cr
height4pt&\omit&&\omit&&\omit&&\omit&&\omit&&\omit&&\omit&\cr
&$V_8$ $(\xi_q = 1/\xi_t$)&&\hfill $425-525$ \hfill&&\hfill TEV \hfill
&&\hfill$4.80$\hfill&&
\hfill$3.30$\hfill&&\hfill $1.50$\hfill&&\hfill $0.37$ \hfill &\cr
height4pt&\omit&&\omit&&\omit&&\omit&&\omit&&\omit&&\omit&\cr
&$V_8$ $(\xi_q = 1/\xi_t$)&&\hfill $425-525$ \hfill&&\hfill LHC \hfill
&&\hfill$159$\hfill&&
\hfill$80$\hfill&&\hfill $178$\hfill&&\hfill $0.012$ \hfill &\cr
height4pt&\omit&&\omit&&\omit&&\omit&&\omit&&\omit&&\omit&\cr}
\hrule\hrule}}
\medskip
\centerline{TABLE 3: Angular dependences of $\ttb$ production in the
$\eta_T$ and $V_8$ resonance models.}

\medskip

\noindent {Top quarks are produced with pseudorapidity $|\eta_{t, \ol t}| <
1.5$ and cross sections (in pb) have been multiplied by 1.62. The $V_8 \ol
t t$ couplig is $\xi_t = \sqrt{40/3}$.}

\vfil\eject

\centerline{\bf Figure Captions}
\bigskip

\item{[1]} The $\tbt$ invariant mass distributions, in $\ppb$ collisions at
$\ecm = 1800\,\gev$, for $m_t = 100-220\,\gev$ in $20\,\gev$ increments.
EHLQ Set 1 distribution functions were used and the cross sections were
multiplied by~1.62 as explained in the text. No rapidity cut is applied.

\medskip

\item{[2]} The $\ttb$ invariant mass distribution in the presence of an
$\eta_T$, in $\ppb$ collisions at $\ecm = 1800\,\gev$, for $m_t =
175\,\gev$ and $\Mh = 450\,\gev$, $F_Q = 30\,\gev$ and $C_t = -1/3$. The
QCD (dotted curve), $\eta_T \ra \ttb$ and its interference with the QCD
amplitude (dashed), and total (solid) rates have been multiplied by 1.62
as explained in the text. No rapidity cut is applied to the top quarks.

\medskip

\item{[3]} The $\ttb$ invariant mass distribution in the presence of an
$\eta_T$, in $\ppb$ collisions at $\ecm = 1800\,\gev$, for $m_t =
175\,\gev$ and $\Mh = 475\,\gev$, $F_Q = 30\,\gev$ and $C_t = -1/3$. The
QCD contribution was multiplied by 1.62 and the $\eta_T$ contribution
by~2.25. Curves are labeled as in Fig.~2.

\medskip

\item{[4]} The $\ttb$ invariant mass distribution in the presence of a
$V_8$, in $\ppb$ collisions at $\ecm = 1800\,\gev$, for $m_t = 175\,\gev$
and $\Mv = 475\,\gev$, $\xi_t = \xi_b = -1/\xi_q = \sqrt{40/3}$. The QCD
(dotted curve) and the total (solid) rates have been multiplied by 1.62 as
explained in the text. No rapidity cut is applied to the top quarks.

\medskip

\item{[5]} The $\ttb$ invariant mass distribution in the presence of a
$V_8$, in $\ppb$ collisions at $\ecm = 1800\,\gev$, for $m_t = 175\,\gev$
and $\Mv = 475\,\gev$, $\xi_t = \xi_b = 1/\xi_q = \sqrt{40/3}$. The curves
are labeled as in Fig.~4.

\medskip

\item{[6]} The $\cstar$ distribution for $\ppb \ra ttb$ at $\ecm =
1800\,\gev$, as defined in the text, in the presence of a 450~GeV $\eta_T$
with parameters as in Fig.~2; $430 < \Mtt < 470\,\gev$. The components are
standard QCD $gg \ra \ttb$ (dot-dash), $\qqb \ra \ttb$ (long dashes), total
QCD (dots), $gg \ra \eta_T \ra \ttb$ and interference with QCD (short
dashes), and the total $d\sigma/\cstar$ (solid). EHLQ Set~1 distribution
functions were used and all cross sections were multiplied by 1.62. The
top quarks are required to have pseudorapidity $|\eta_{t, \ol t}| < 1.5$.

\medskip

\item{[7]} The $\cstar$ distribution for $pp \ra \ttb$ at $\ecm =
15\,\tev$, as defined in the text, in the presence of a 450~GeV $\eta_T$
with parameters as in Fig.~2; $430 < \Mtt < 470\,\gev$. The components are
labeled as in Fig.~6. The top quarks are required to have pseudorapidity
$|\eta_{t, \ol t}| < 1.5$.

\medskip

\item{[8]} The $\cstar$ distribution for $\ppb \ra \ttb$ at $\ecm =
1800\,\gev$ in the presence of a 475~GeV $V_8$ with parameters as in
Fig.~4; $400 < \Mtt < 500\,\gev$. The components are standard QCD $gg \ra
\ttb$ (dot-dash), $\qqb \ra \ttb$ (long dashes), total QCD (dots), $\qqb
\ra V_8 \ra \ttb$ and interference with QCD (short dashes), and the total
$d\sigma/\cstar$ (solid). EHLQ Set~1 distribution functions were used and
all cross sections were multiplied by 1.62. The top quarks are required to
have pseudorapidity $|\eta_{t, \ol t}| < 1.5$.

\medskip

\item{[9]} The $\cstar$ distribution for $pp \ra \ttb$ at $\ecm = 15\,\tev$
in the presence of a 475~GeV $V_8$ with parameters as in Fig.~4; $400 <
\Mtt < 500\,\gev$. The components labeled as in Fig.~8. The top quarks are
required to have pseudorapidity $|\eta_{t, \ol t}| < 1.5$.

\medskip

\item{[10]} The $\cstar$ distribution for $pp \ra \ttb$ at $\ecm = 15\,\tev$
in the presence of a 475~GeV $V_8$ with parameters as in Fig.~4; $400 <
\Mtt < 500\,\gev$. The components labeled as in Fig.~8. The top quarks are
required to have pseudorapidity $|\eta_{t, \ol t}| < 2.5$.

\vfil\eject

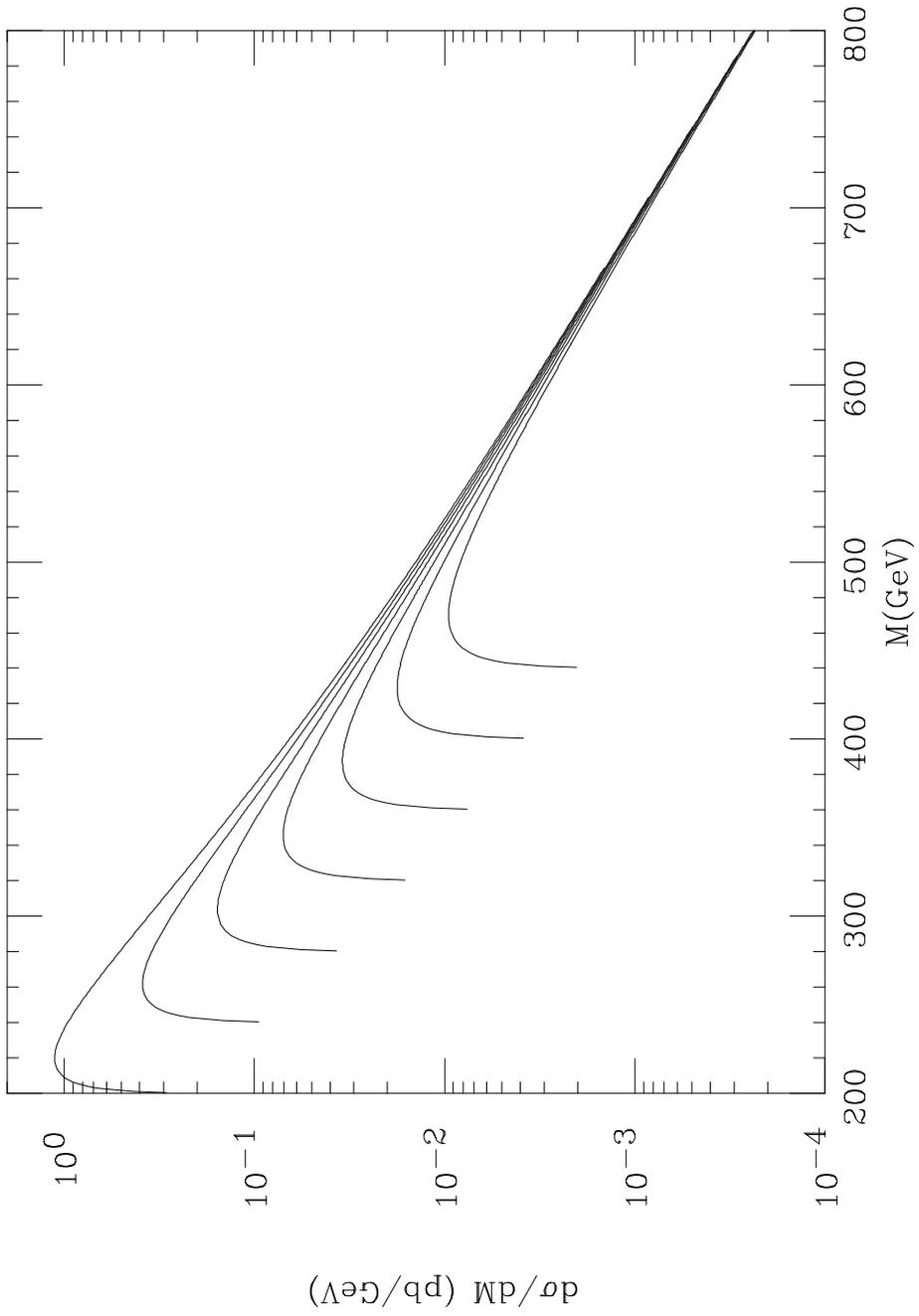
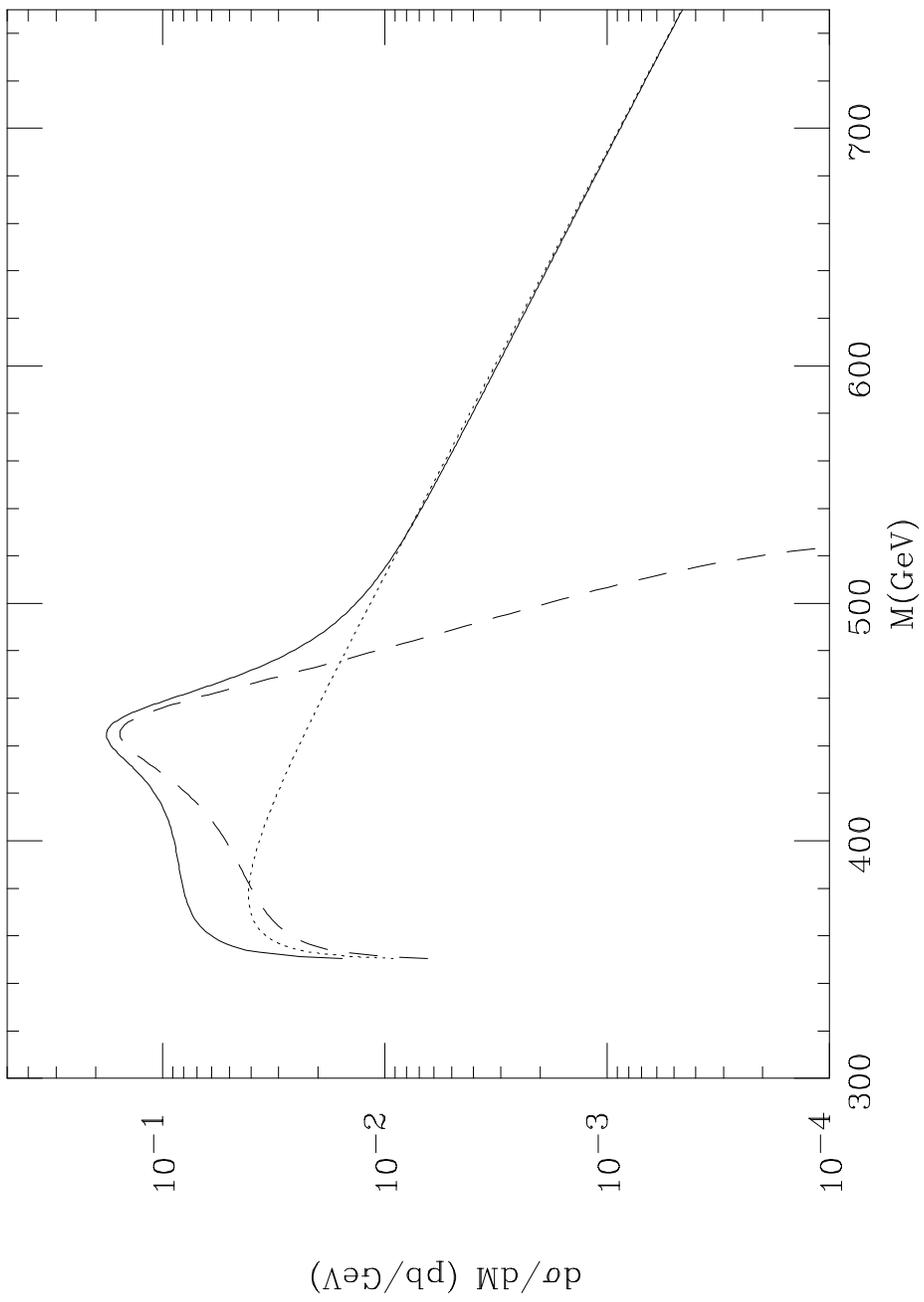
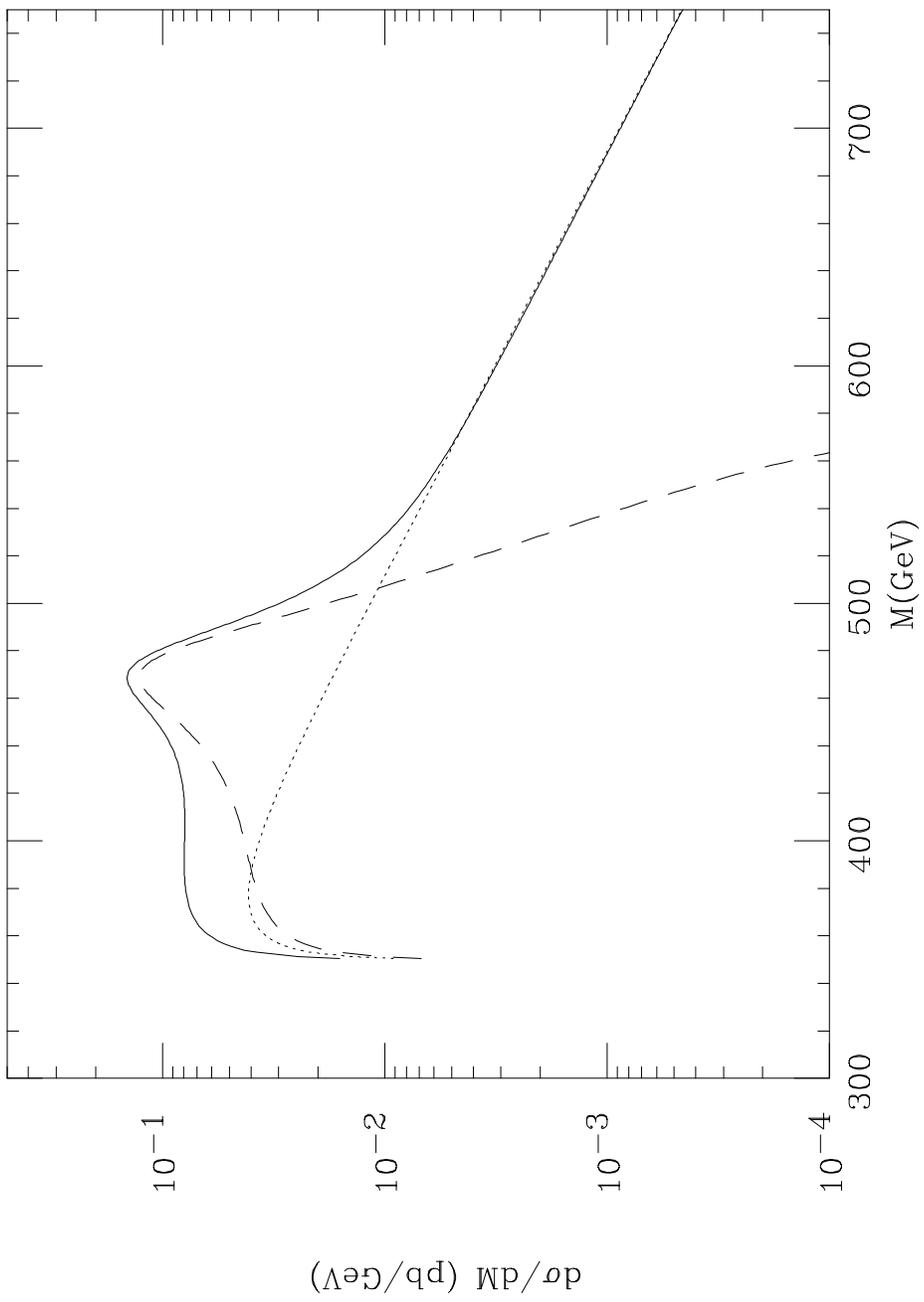
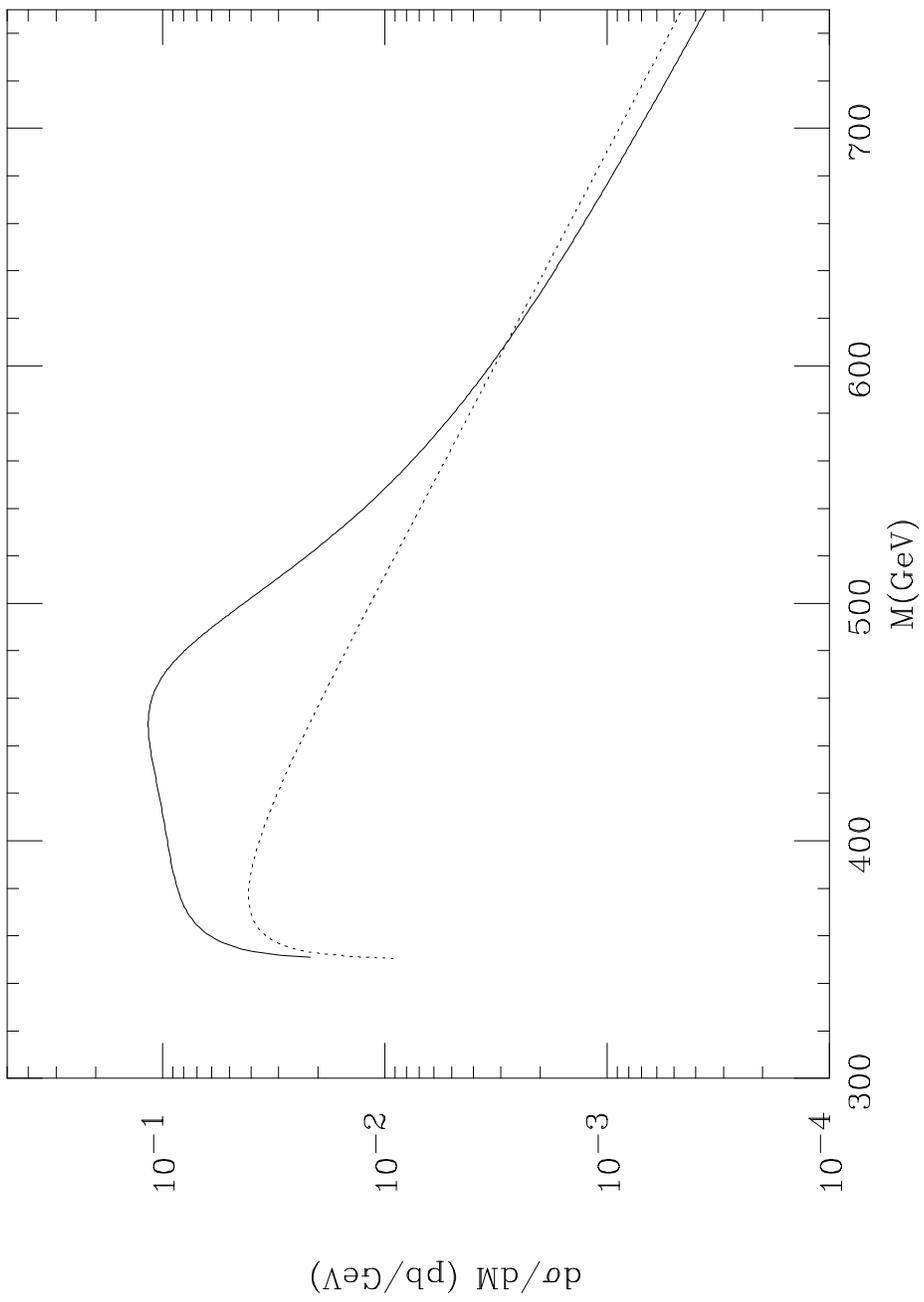
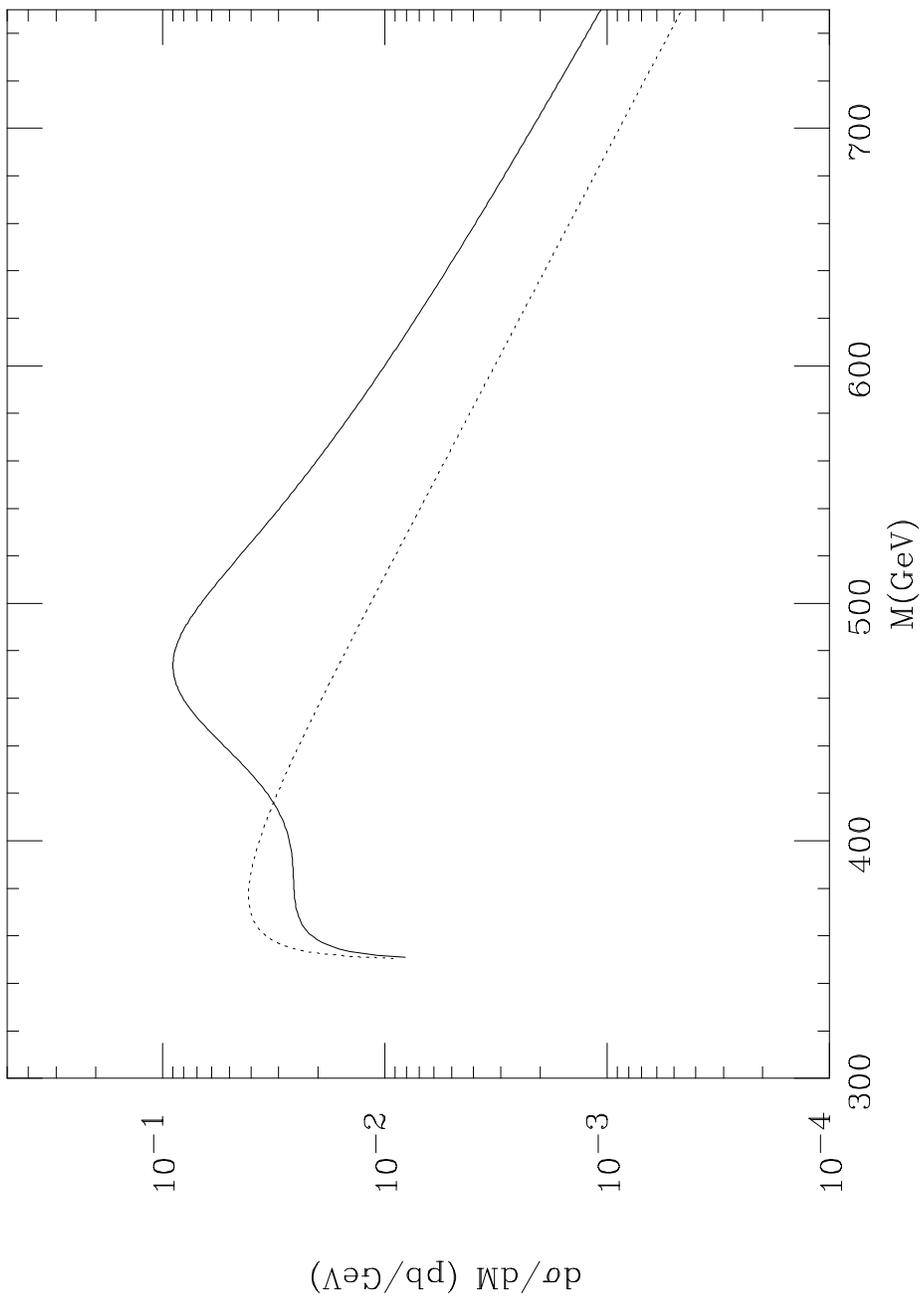
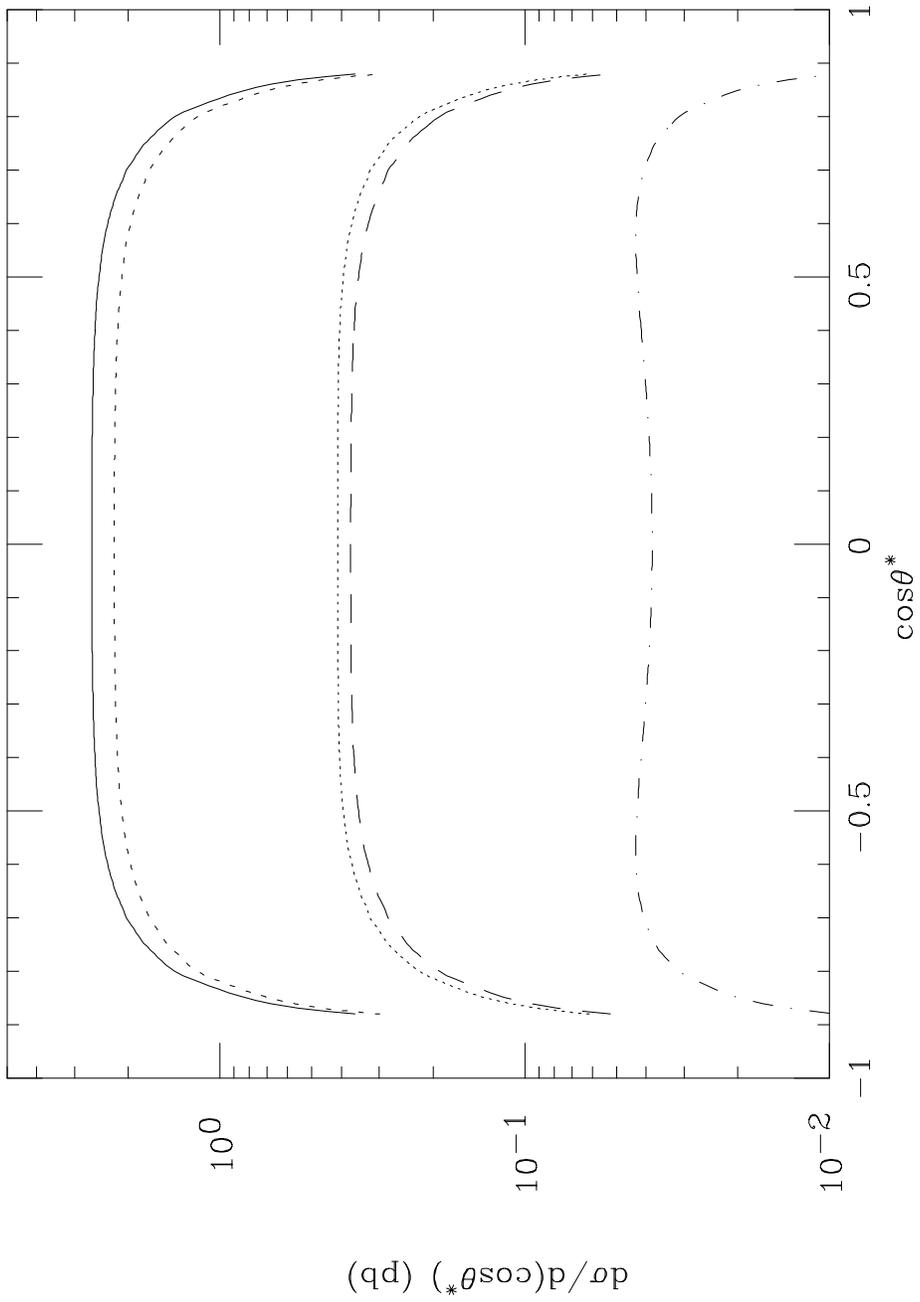
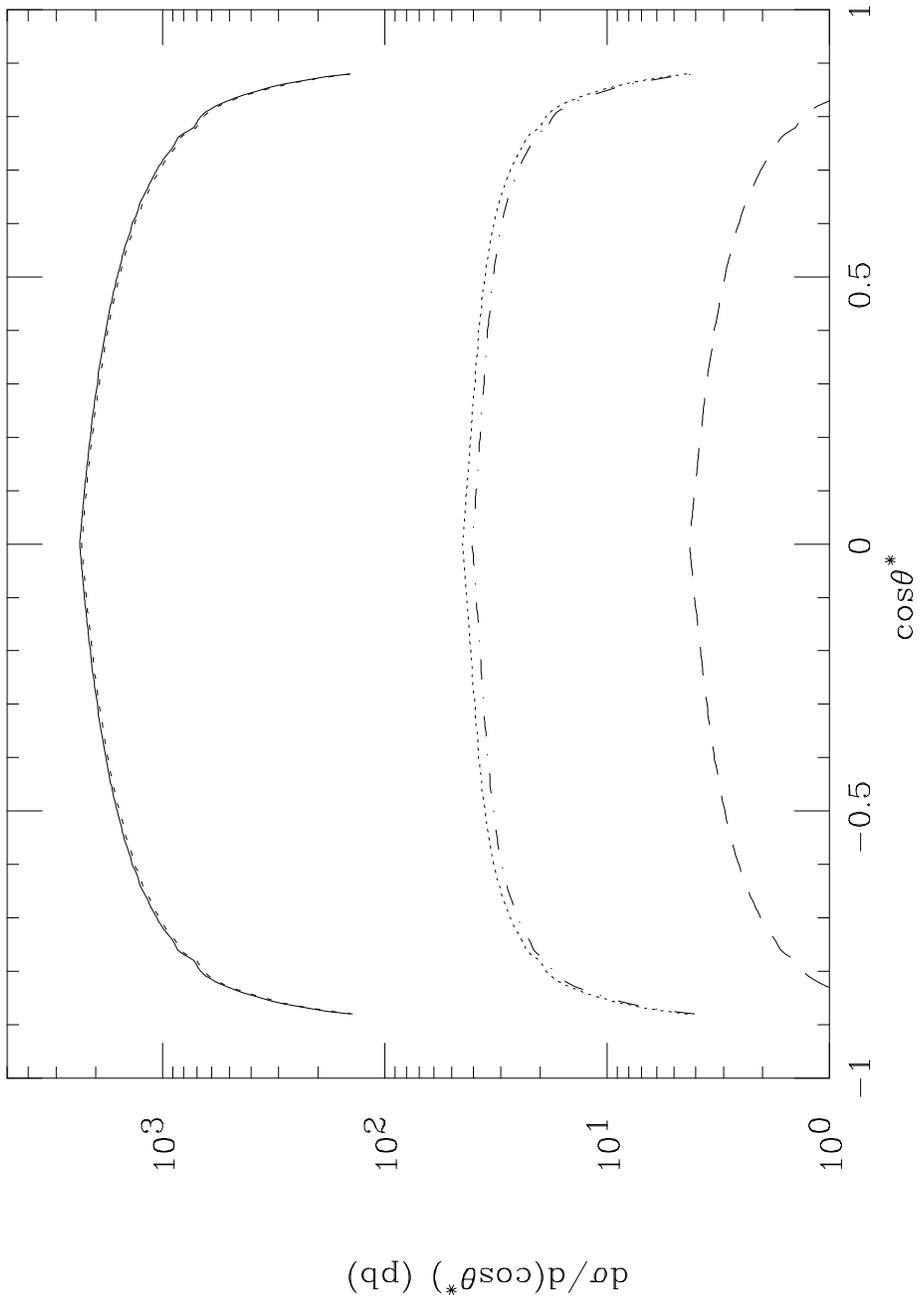
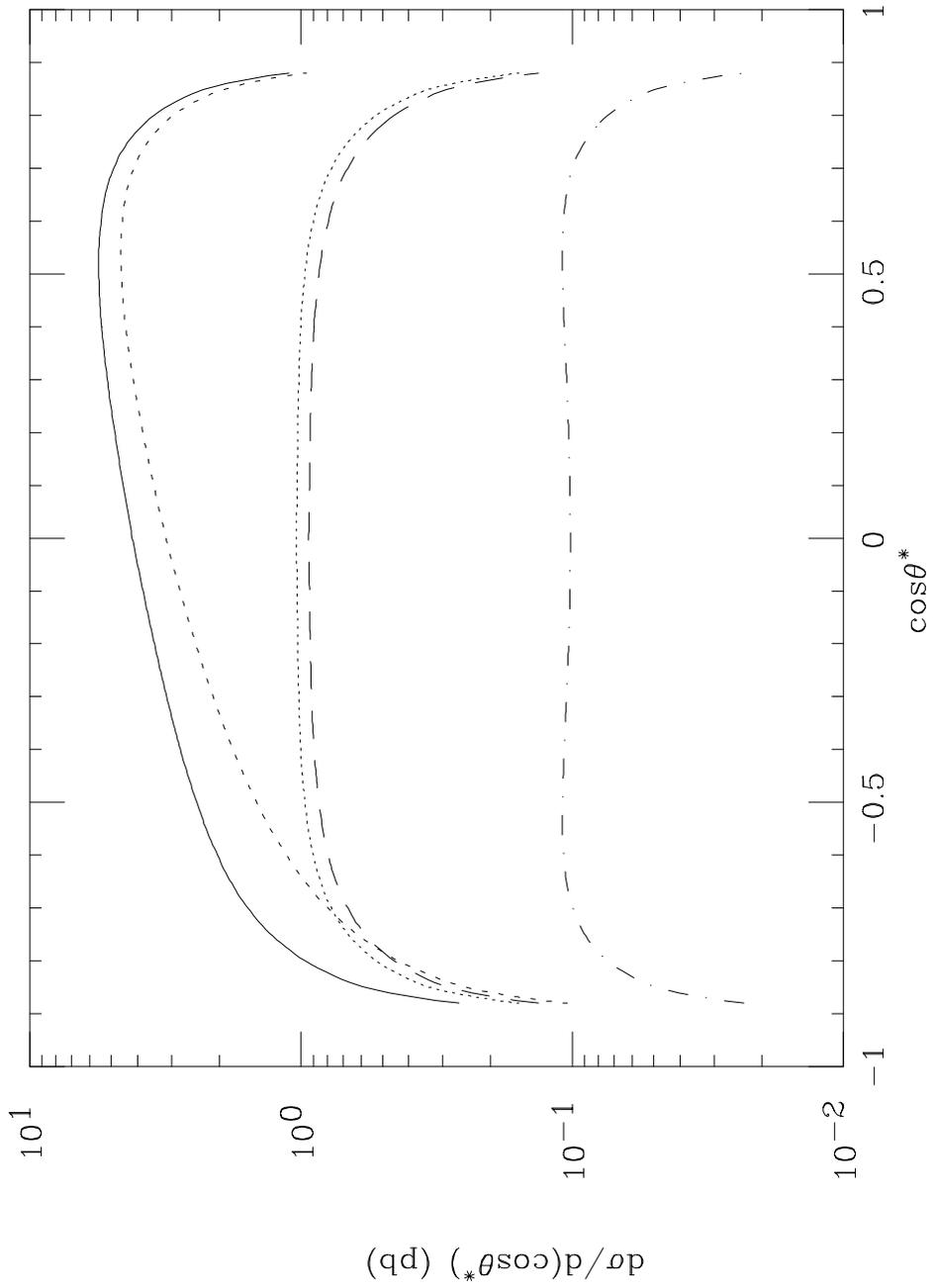
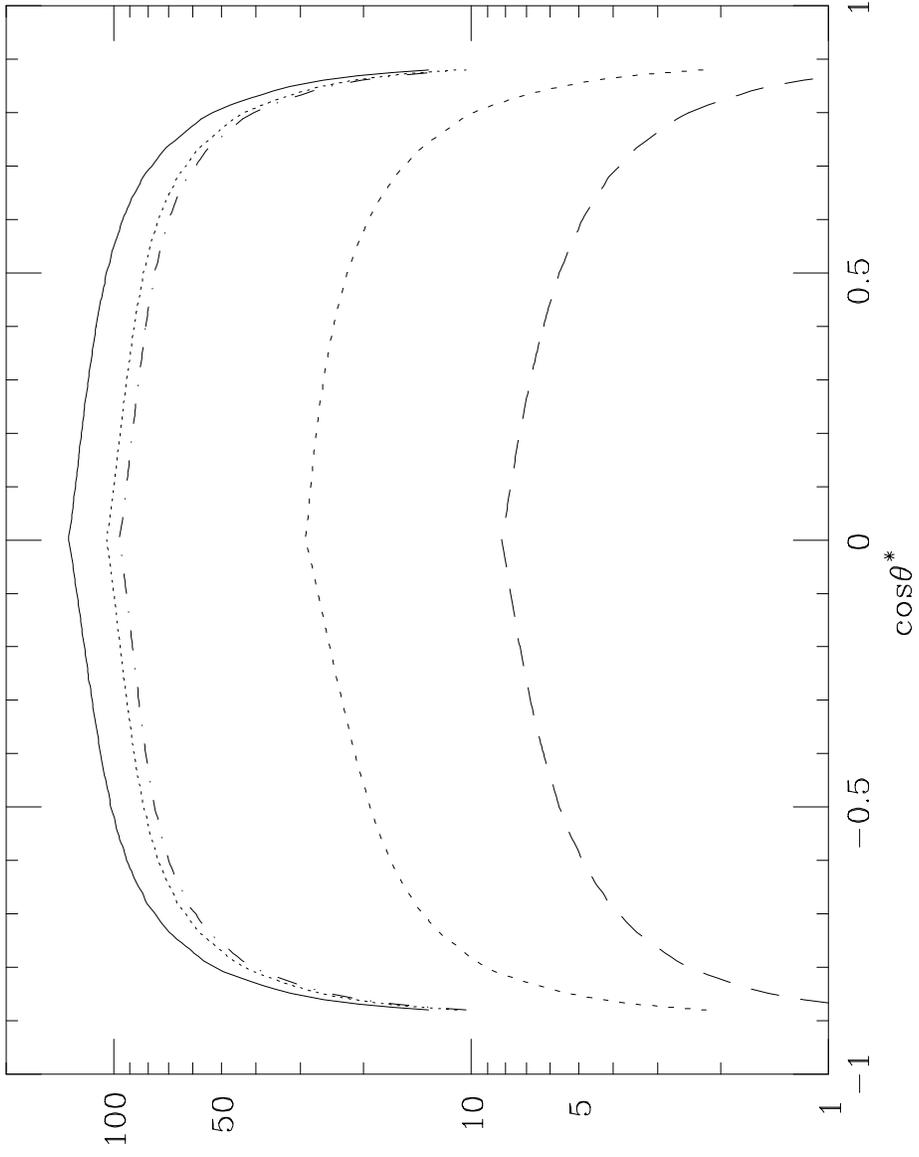
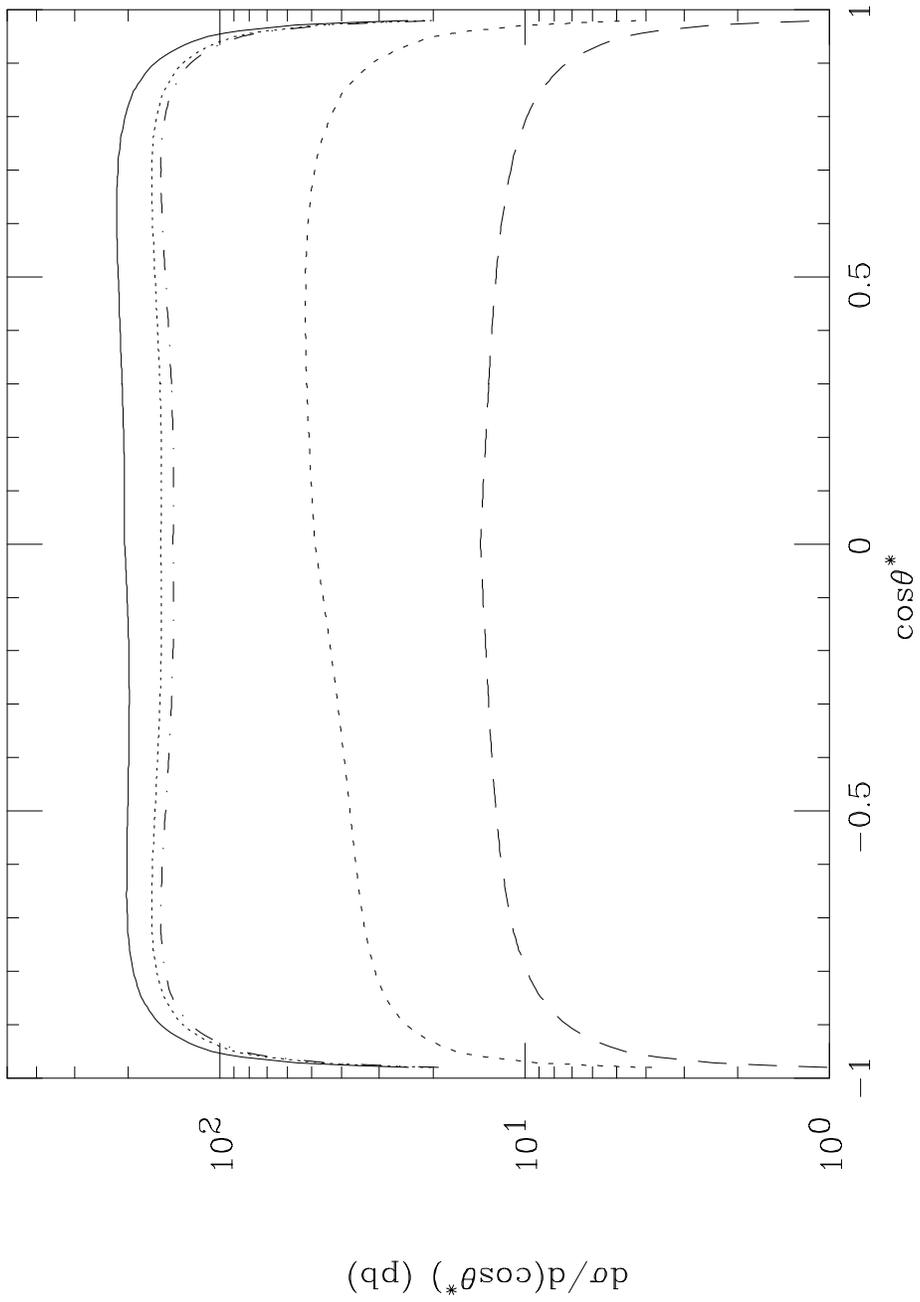
\bye